\def \be {\begin{equation}}
\def \eq {\end{equation}}
\def \bee {\begin{eqnarray}}
\def \eqq {\end{eqnarray}}

\def \bea {\begin{array}{c}}
\def \eqa {\end{array}}

\def \la {\left\langle}
\def \ra {\right\rangle}

\def \dels {\partial\kern-.5em / \kern.5em}
\def \two {I\hspace{-.1em}I\hspace{.2em}}

\documentstyle[12pt]{article}
\input epsf
\def\Tr{{\rm Tr}}
\newcommand{\Lag}{{\cal L}}
\newcommand{\superint}{\int \diff^{4}\theta}
\newcommand{\lowest}{|_{\theta =\bar{\theta}=0}}
\newcommand{\diff}{\mbox{d}}
\newcommand{\Diff}{{\cal D}}

\def\[{\left [}
\def\]{\right ]}
\def\({\left (}
\def\){\right )}
\def\lbr{\left\{}
\def\rbr{\right\}}
\def\pp{\partial}
\def\CO{{\cal O}}
\def\l{\ell}
\def\tl{\tilde{\ell}}

\def\df{f_{\ell}}
\def\tf{\tilde{f}}
\def\dtf{\tilde{f}_{\tilde{\ell}}}

\def\dg{g_{\ell}}
\def\tg{\tilde{g}}
\def\dtg{\tilde{g}_{\tilde{\ell}}}
\def\tA{\tilde{A}}
\def\tB{\tilde{B}}
\def\tN{\tilde{N}}
\def\tV{\tilde{V}}
\def\tD{\tilde{D}}
\def\tG{\tilde{G}}
\def\tS{\tilde{S}}
\def\ts{\tilde{s}}
\def\ta{\tilde{a}}

\def\tF{\tilde{{\cal F}}}
\def\cA{{\cal P}}
\def\cB{{\cal Q}}
\def\cV{{\cal V}}
\def\tcA{\tilde{{\cal P}}}
\def\tcB{\tilde{{\cal Q}}}
\def\ls{linear supermultiplet }

\def\cs{chiral supermultiplet }

\def\Kahler{K\"{a}hler }

\renewcommand{\theequation}{\arabic{section}.\arabic{equation}}
\begin{document}
\begin{titlepage}
\begin{center}

\vskip 3.0in
{\hfill $\;$}
{\hfill $\;$}

{\large \bf Gaugino Condensation in ${\cal N}$=1 Supergravity
            \\ \vspace{0.2cm}
            Models with Multiple Dilaton-Like Fields}

\vskip .4in

{\large Jonathan Bagger}\footnote{E-mail: {\tt bagger@jhu.edu}}
and {\large Yi-Yen Wu}\footnote{E-mail: {\tt yywu@eta.pha.jhu.edu}}

\vskip .2in
{\em Department of Physics and Astronomy \\
           Johns Hopkins University \\
          Baltimore, Maryland 21218, USA}

\end{center}

\vskip 0.3in
\begin{abstract}
We study supersymmetry breaking by hidden-sector
gaugino condensation in ${\cal N}$=1 D=4 supergravity
models with multiple dilaton-like moduli fields.
Our work is motivated by Type I string theory, in
which the low-energy effective Lagrangian can have
different dilaton-like fields coupling to different
sectors of the theory.  We construct the effective
Lagrangian for gaugino condensation and use it
to compute the visible-sector gaugino masses.  We
find that the gaugino masses can be of order the
gravitino mass, in stark contrast to heterotic
string models with a single dilaton field.
\end{abstract}
\vskip .0in

\end{titlepage}

\newpage
\renewcommand{\thepage}{\arabic{page}}
\setcounter{page}{1}
\setcounter{footnote}{0}

\section{Introduction} \label{intro}
\setcounter{equation}{0}

Before the recent string duality revolution,
most string phenomenology centered on perturbative
${\cal N}$=1 D=4 heterotic string theories.
The discovery of string duality and D-branes,
however, opened a variety of new approaches to
string phenomenology based on Type I and Type
\two string theories.  For example, much recent
work has focussed on the intriguing possibility
that our four-dimensional world lies at the
intersection of a set of D$p$-branes (3$\le$
$p$$\le$9) embedded in $9+1$-dimensional
spacetime.\footnote{See
\cite{Ibanez1}\cite{Ibanez2}\cite{Tye1}
for recent reviews.}

A well-known problem with the usual perturbative
heterotic string phenomenology is that hidden-sector
gaugino condensation \cite{Rohm} gives rise to
visible-sector gaugino masses that are much smaller
than the scale of supersymmetry breaking
\cite{moduli}\cite{versus}. This is a consequence of
the fact that a single dilaton couples to all gauge
and matter fields. Gaugino condensation induces a small
$F$ term for the dilaton field, and the dilaton
couplings then give small masses to the gauginos.

In Type I models, however, the situation can be
very different.  In these models, the hidden and
visible sectors can live on different D-branes.
Each sector has its own dilaton-like fields
\cite{Ibanez2}. The hidden-sector dilatons receive
small $F$ terms from gaugino condensation. However,
these $F$ terms are not responsible for the
visible-sector gaugino masses, and therefore the
gaugino masses are not forced to be small.

Inspired by this possibility, in this paper we study
the question of supersymmetry breaking by hidden-sector
gaugino condensation in ${\cal N}$=1 D=4 supergravity
models with multiple dilaton-like fields. We compute the
visible-sector gaugino masses and find that
they can indeed be of order the supersymmetry
breaking scale.  We see that Type I models
offer an appealing solution to the gaugino
mass problem associated with heterotic
string theories.

We approach this problem in the spirit of
effective field theory.  We take our visible
sector to be composed of multiple pure ${\cal
N}$=1 super Yang-Mills theories, each coupled
to gravity, and each to its own dilaton-like field.
We take our hidden sector to be composed of
gaugino condensate fields, one on each set of
branes. The condensate fields are also coupled
to gravity, and to their associated
dilatons.\footnote{To simplify our presentation,
we ignore all charged chiral
superfields; this restriction does not affect
the results of our analysis.}  By construction,
this theory describes the low-energy limit of
a Type I string theory, where each super Yang-Mills
theory lives on its own set of D-branes.  Of
course, our analysis also applies to other
string/M theory vacua with multiple dilaton-like
fields.

Various ${\cal N}$=1 D=4 Type I models have been
proposed in the literature \cite{Berkooz}--\cite{Tye2}.
Particularly simple examples can be constructed
from Type I models with D9-branes and D5-branes
compactified on $T^{2}\!\times\! T^{2}\!\times\!
T^{2}$, where $R_{i}$ is the radius of the $i$th
two-torus, $i=1,2,3$.  (D5-branes that wrap on
the $i$th two-torus are denoted as D5$_{i}$-branes.)
These models have one dilaton field $S$, three
untwisted moduli fields $T_{i}$ ($i=1,2,3$) and
additional twisted moduli fields from the closed
string sector. Other D-brane configurations
can be obtained from these by T-duality.

One such example is a Type I model with two
sectors, built from D9-branes and D5$_{1}$-branes.
The gauge bosons arise from open strings ending
on the D9- and D5$_{1}$-branes. The low-energy
effective Lagrangian is as follows\footnote{For
convenience, we use the K\"{a}hler superspace
formulation throughout this paper \cite{Binetruy90}.}
\cite{Ibanez1}\cite{Ibanez4},
\begin{eqnarray} \label{intro1}
\Lag\,&=&\,\hspace{0.42cm}
\frac{1}{8}\superint\,\frac{E}{R}\,
S({\cal W}^{\alpha}{\cal W}_{\alpha})_{D9}\,+\,
\frac{1}{8}\superint\,\frac{E}{R^{\dagger}}\,\bar{S}
({\cal W}_{\dot{\alpha}}{\cal W}^{\dot{\alpha}})_{D9} \nonumber \\
\,& &\,
\,+\,\frac{1}{8}\superint\,\frac{E}{R}\,
T_{1}({\cal W}^{\alpha}{\cal W}_{\alpha})_{D5_{1}}
\,+\,\frac{1}{8}\superint\,\frac{E}{R^{\dagger}}\,\bar{T}_{1}
({\cal W}_{\dot{\alpha}}{\cal W}^{\dot{\alpha}})_{D5_{1}}\;+\;\cdots
\end{eqnarray}
where
\begin{equation}
K\,=\,-\ln\(S+\bar{S}\)\,
-\,\sum_{i=1}^{3}\ln\(T_{i}+\bar{T}_{i}\)\;+\;\cdots
\end{equation}
is the \Kahler function.
The dots denote possible contributions from other
neutral chiral superfields, and
\begin{equation} \label{intro2}
\la S+\bar{S}\ra\,=\,
\frac{R_{1}^{2}R_{2}^{2}R_{3}^{2}}{\pi\lambda_{I}\alpha'^{3}},
\;\;\;\;\;
\la T_{i}+\bar{T}_{i}\ra\,=\,
\frac{R_{i}^{2}}{\pi\lambda_{I}\alpha'},\;\;\;\;\;i=1,2,3.
\end{equation}
\vskip 0.1in
\noindent
In these expressions, $\lambda_{I}$ is the string coupling,
$\alpha'=M_{I}^{-2}$, and $M_{I}$ is the string scale. The
field strengths $({\cal W}^{\alpha} {\cal W}_{\alpha})_{
D9}$ and $({\cal W}^{\alpha}{\cal W}_{\alpha})_{D5_{1}}$
contain gauge fields living on D9- and D5$_{1}$-branes,
respectively.  Note that the moduli $S$, $T_{1}$ are
dilaton-like fields, while $T_{2}$, $T_{3}$ are
not.\footnote{Our definition of the $S$ and $T_{i}$
moduli agrees with \cite{Binetruy90}, but
differs from \cite{Ibanez1}.}

A second example contains D9-branes as well as D5$_{i}$-branes
compactified on all three tori ($i = 1,2,3$). It has four
sectors, and four dilaton-like fields, $S$, $T_{1}$, $T_{2}$,
$T_{3}$, as defined in (\ref{intro2}). These fields couple
to the gauge fields on the D9- and D5$_{i}$-branes
and give rise to the effective Lagrangian
\begin{eqnarray} \label{intro3}
\Lag&=&\!\!\!\!\!\!\!\!\!\!\!\!\hspace{1.1cm}
\frac{1}{8}\superint\,\frac{E}{R}\,
S({\cal W}^{\alpha}{\cal W}_{\alpha})_{D9}\,+\,
\frac{1}{8}\superint\,\frac{E}{R^{\dagger}}\,\bar{S}
({\cal W}_{\dot{\alpha}}{\cal W}^{\dot{\alpha}})_{D9}
\nonumber \\
& &\!\!\!\!\!\!\!\!\!\!\!\!
\,+\,\sum_{i=1}^{3}
\frac{1}{8}\superint\,\frac{E}{R}\,
T_{i}({\cal W}^{\alpha}{\cal W}_{\alpha})_{D5_{i}}
\,+\,\sum_{i=1}^{3}\frac{1}{8}\superint\,\frac{E}{R^{\dagger}}\,
\bar{T}_{i}
({\cal W}_{\dot{\alpha}}{\cal W}^{\dot{\alpha}})_{D5_{i}}\;+\;\cdots
\end{eqnarray}

The plan of this paper is as follows.  In Section
\ref{model}, we define our ${\cal N}$=1 D=4
supergravity model with multiple dilaton-like fields.
In Section \ref{gaugino}, we argue that the extra
dilaton-like fields allow gaugino masses to be
as large as the gravitino mass. We also present
an explicit example of this scenario.

In Appendix A, we exhibit the linear--chiral duality
for ${\cal N}$=1 D=4 supergravity models with multiple
linear supermultiplets. In Appendix B we extract
the relevant pieces of the component supergravity
Lagrangian. Finally, in Appendix C we check our
results by comparing with the
\cs formulation.

\section{Supergravity with Multiple Dilaton-Like Fields}
\label{model} \setcounter{equation}{0}

In this section we define the ${\cal N}$=1
D=4 supergravity model that we will study.
We start by considering a system with
$N$ different types of D-branes.  Of these,
we take $\tN$ to be D$p_{\tA}$-branes
($\tA=1,\cdots,\tN$) with weakly coupled visible-sector
fields on their world volumes. We take the remaining
$N\!-\!\tN$ to be D$p_{A}$-branes ($A=1,\cdots,N-\tN$)
with strongly coupled hidden-sector fields in the
condensation phase.

Let us first construct the effective theory
of the visible sector. For each value of $\tA$,
let ${\cal W}_{\tA}$ denote the super Yang-Mills
field strength on the $\tA$-th set of branes, and
let $\tV_{\tA}$ be the real superfield which
contains the $\tA$-th dilaton-like field
\cite{Linear}. The field $\tV_{\tA}$
obeys the constraint
\begin{equation} \label{model1}
-(\bar{\Diff}^{2}-8R)\tV_{\tA}\,=\,
{\cal W}_{\tA}{\cal W}_{\tA}.
\end{equation}
This constraint couples the dilatons to the
super Yang-Mills fields.

In \Kahler superspace \cite{Binetruy90}, the supergravity
kinetic terms
are given by a Lagrangian $\Lag$ and a \Kahler
function $K$. For the visible-sector fields, we
take them to be
\begin{eqnarray} \label{model5}
\Lag\,&=&\,\superint\,E\,\lbr\; -3\,+\,\tN\,+\,
\sum_{\tA=1}^{\tN}\tf_{\tA}(\tV_{\tA})\;\rbr, \nonumber\\
K\,&=&\,\sum_{\tA=1}^{\tN}\lbr\;
\ln {\tV}_{\tA}\,+\,\tg_{\tA}(\tV_{\tA})\;\rbr,
\end{eqnarray}
where
\begin{equation} \label{model6}
\tV_{\tA}\frac{\diff \tg_{\tA}}{\diff \tV_{\tA}} \,=
\,\tf_{\tA}\,-\,\tV_{\tA}\frac{\diff \tf_{\tA}}{\diff
\tV_{\tA}}.
\end{equation}
In these expressions, the
leading terms describe the tree-level couplings of
the gauge and dilaton-like fields; the functions
$\tg_{\tA}(\tV_{\tA})$ and $\tf_{\tA}(\tV_{\tA})$
contain corrections beyond tree level. The condition
(\ref{model6}) guarantees that the Einstein gravity
term is canonically normalized. (For simplicity, we
do not include mixings between the different
$\tV_{\tA}$'s. Such mixings arise at the loop level,
but they do not change our conclusions.\footnote{Type
I models typically contain chiral superfields charged
under the D$p_{\tA}$ and D$p_{\tA'}$ gauge groups
($\tA\neq \tA'$).  These superfields mix the different
dilaton-like fields. Our study ignores charged chiral
superfields, so it is consistent to ignore mixings
between different $\tV_{\tA}$'s.})

In the hidden sector, the effective Lagrangian is different
because the super Yang-Mills fields are in a strongly-coupled
condensation phase. This leads us to replace the
Yang-Mills fields by
chiral condensate superfields $U_{A}$ \cite{Vene}. These
fields are contained within the real superfields $V_{A}$
\cite{Burgess95}\cite{Binetruy95}, where
\begin{equation}
-(\bar{\Diff}^{2}-8R) V_{A}\,=\,
U_{A}.
\end{equation}
The hidden-sector dilaton-like fields are also contained
in the fields $V_{A}$.

The effective Lagrangian for this sector contains
kinetic and superpotential terms.  The kinetic
terms are as above,
\begin{eqnarray}
\Lag\,&=&\,\superint\,E\,\lbr\;-3\,+\,N-\tN\,+\,
\sum_{A=1}^{N-\tN}f_{A}(V_{A})\;\rbr, \nonumber\\
K\,&=&\,\sum_{A=1}^{N-\tN}\lbr\;
\ln {V}_{A}\,+\,g_{A}(V_{A})\;\rbr,
\end{eqnarray}
where
\begin{equation}
V_{A}\frac{\diff g_{A}}{\diff V_{A}} \,=
\,f_{A}\,-\,V_{A}\frac{\diff f_{A}}{\diff V_{A}}.
\end{equation}
The leading terms describe the tree-level
couplings, while $g_{A}(V_{A})$ and $f_{A}(V_{A})$
contain corrections beyond tree level.

The superpotential terms are generated by nonperturbative
effects associated with gaugino condensation. In what
follows, we take the superpotential to be given by
\begin{eqnarray} \label{cond1}
\Lag_{A}\,&=&\,
\superint\,\frac{E}{R}\,
\frac{1}{8}b_{A}U_{A}\ln(e^{\!-K/2}U_{A})\,+\,\mbox{h.c.}
\nonumber\\[2mm]
&=&\,\superint\,E\,\,b_{A}V_{A}\ln\(e^{\!-K}\bar{U}_{A}U_{A}\),
\end{eqnarray}
where $b_{A}$=$2b'_{A}/3$, and $b'_{A}$ is the one-loop
$\beta$-function coefficient of D$p_{A}$-sector. The
form of this term is dictated by the anomalies
of the underlying super Yang-Mills theory
\cite{Rohm}\cite{Vene}\cite{chiral91}\cite{KL}\cite{DB}.

In rigid supersymmetry, the superpotential
is fixed by the chiral and conformal anomalies \cite{Vene}.
In local supersymmetry, the \Kahler anomaly also comes
into play \cite{Ferrara}\cite{Ovrut}. In particular,
it fixes
the explicit $K$ dependence of the superpotential.
Under an arbitrary \Kahler
transformation, $K\rightarrow K+ F+\bar{F}$, the fields
$V_{A} \rightarrow V_{A}$ and
$U_{A}\rightarrow U_{A} e^{(\bar{F}-F)/2}$.
The superpotential then transforms as follows,
\begin{eqnarray}
\Lag_{A}\,& \equiv &\,\superint\,E\;
b_{A}V_{A}\ln(e^{-K}\bar{U}_{A}U_{A})\nonumber \\
& \rightarrow &\,\Lag_{A}
\,-\,\superint\,\frac{E}{R}\,\frac{1}{8}b_{A}U_{A}F\,-\,
\superint\,\frac{E}{R^{\dagger}}
\,\frac{1}{8}b_{A}\bar{U}_{A}\bar{F}.
\end{eqnarray}
This is precisely the right transformation to match the
\Kahler anomaly of the underlying super Yang-Mills theory
\cite{Rohm}\cite{chiral91}\cite{KL}\cite{DB}.

The full supergravity Lagrangian contains contributions
from both of these sectors. It can be written as
follows,\footnote{Note also that these expressions contain
contributions from extra moduli fields $\Phi_{i}$,
$i=1,\cdots,n$. In the context of Type I models, the
$\Phi_{i}$ are twisted or non-dilaton-like untwisted
moduli fields.}
\begin{eqnarray}
\Lag\,&=&\,\superint\,E\,\lbr\;-3\,+\,N\,+\,
\sum_{\tA=1}^{\tN}\tf_{\tA}(\tV_{\tA})
\,+\,\sum_{A=1}^{N\!-\!\tN}f_{A}(V_{A})\;\rbr \nonumber \\[2mm]
& &\!\!\!+\,\sum_{A=1}^{N\!-\!\tN}\;
\superint\,E\,\,b_{A}V_{A}\ln\(e^{\!-K}\bar{U}_{A}U_{A}\),
\label{cond3}
\end{eqnarray}
where the \Kahler function is given by
\begin{eqnarray}
K\,&=&\,\sum_{\tA=1}^{\tN}\lbr\;\ln \tV_{\tA}\,+\,
\tg_{\tA}(\tV_{\tA})\;\rbr\,+\,
\sum_{A=1}^{N\!-\!\tN}\lbr\;\ln V_{A}\,+\,g_{A}(V_{A})\;\rbr
\nonumber \\[3mm]
& &\!\!\!+\,G(\Phi_{1},\bar{\Phi}_{1},\cdots,\Phi_{n},
\bar{\Phi}_{n}),
 \label{cond2}
\end{eqnarray}
and
\begin{eqnarray}
\tV_{\tA}\frac{\diff \tg_{\tA}}{\diff \tV_{\tA}}\,&=&\,\tf_{\tA}
\,-\,\tV_{\tA}\frac{\diff \tf_{\tA}}{\diff \tV_{\tA}},
\hspace{2.0cm}\tA=1,\cdots,\tN, \label{cond4}\nonumber\\
V_{A}\frac{\diff g_{A}}{\diff V_{A}}\,&=&\,f_{A}\,
-\,V_{A}\frac{\diff f_{A}}{\diff V_{A}},
\hspace{2.0cm}A=1,\cdots,N\!-\!\tN. \label{cond5}
\end{eqnarray}
The hidden-sector superpotential couples all
sectors through its explicit $K$ dependence; it
gives rise to gaugino
masses in the visible sector.

\section{Gaugino Masses} \label{gaugino} \setcounter{equation}{0}

To find the gaugino masses, we need certain terms
from the component-field Lagrangian. These terms
are computed in Appendix B.

{}From (\ref{cond29}) we find that the scalar potential takes
the following form,
\begin{eqnarray} \label{cond40}
\cV\!\!
&=&\!\!\sum_{A=1}^{N\!-\!\tN}
\frac{1}{16\l_{A}^{2}}\(1+f^{A}-\l_{A}\df^{A}\)\bar{u}_{A}u_{A}
\nonumber\\
& &\!\!+\,\frac{1}{16}\lbr\,
\(\,\sum_{A=1}^{N\!-\!\tN}
\frac{\(1+f^{A}-\l_{A}\df^{A}\)}{\l_{A}}u_{A}\,\)
\!\!\(\sum_{B=1}^{N\!-\!\tN}b_{B}\bar{u}_{B}\)
\,+\,\mbox{h.c.}\,\rbr
\nonumber\\
& &\!\!+\,\frac{1}{16}
\left\{\begin{array}{lll}&\vspace{0.1cm}
\,-\,3\,\,+\,
\sum_{i,j}G_{\bar{j}}G^{-1}_{\bar{j}i}G_{i} \vspace{0.23cm}&\\
& +\,\sum_{\tA=1}^{\tN}\(1+\tf^{\tA}-\tl_{\tA}\dtf^{\tA}\)
\,+\,\sum_{A=1}^{N\!-\!\tN}\(1+f^{A}-\l_{A}\df^{A}\)
\vspace{0.1cm}& \end{array}\right\}
\!\left|\sum_{A=1}^{N\!-\!\tN}b_{A}\bar{u}_{A}\right|^{2}.
\nonumber\\
& & \!\!\!\!\!\!\!\!\!\!\!\!
\end{eqnarray}
In this expression, $\tl_{\tA}\,$=$\,\tV_{\tA}\lowest\,$
and $\,\l_{A}\,$=$\,V_{A}\lowest\,$ are the dilaton-like
scalar fields which couple to the visible and hidden
sectors, respectively.  Moreover,
$u_{A}\,$=$\,U_{A}\lowest$ is the gaugino condensate field
on each of the D$p_{A}$-branes; it depends on $\l_{A}$
according to (\ref{cond34}). Other notation
is as follows,
\begin{eqnarray} \label{cond27}
\dg^{A}\,&=&\,\frac{\diff g_{A}(V_{A})}{\diff V_{A}}\lowest,
\;\;\;\;\;
\dtg^{\tA}\,=\,
\frac{\diff \tg_{\tA}(\tV_{\tA})}{\diff \tV_{\tA}}\lowest,\nonumber\\
\df^{A}\,&=&\,\frac{\diff f_{A}(V_{A})}{\diff V_{A}}\lowest,
\;\;\;\;\;
\dtf^{\tA}\,=\,
\frac{\diff \tf_{\tA}(\tV_{\tA})}{\diff \tV_{\tA}}\lowest.\nonumber
\end{eqnarray}\vspace{-0.4cm}
\begin{equation} \label{cond28}
G_{i}=\frac{\pp G}{\pp \Phi_{i}}\lowest, \;\;\;\;\;
G_{\bar{j}}=\frac{\pp G}{\pp \bar{\Phi}_{j}}\lowest, \;\;\;\;\;
G_{i\bar{j}}=\frac{\pp^{2} G}{\pp \Phi_{i} \pp \bar{\Phi}_{j}}\lowest.
\end{equation}

The gaugino mass has its origin in the explicit $K$ dependence
of the hidden-sector superpotential. We find (\ref{cond30})
\begin{eqnarray}
m^{\tA}_{\lambda}\,&=&\,\la\;
\frac{1+\tf^{\tA}-\tl_{\tA}\dtf^{\tA}}{1+\tf^{\tA}}
\cdot\sum_{A=1}^{N\!-\!\tN}\;\frac{1}{4}\,b_{A}u_{A} \;\ra,
\label{cond42} \nonumber\\
&=&\,\la\;\frac{1+\tf^{\tA}-\tl_{\tA}\dtf^{\tA}}{1+\tf^{\tA}}\;\ra
m_{\tilde{G}},
\hspace{1.5cm} \tA=1,\cdots,\tN.
\label{cond43}
\end{eqnarray}
The gravitino mass is simply
\begin{equation} \label{cond41}
m_{\tilde{G}}\,=\,\la\;\sum_{A=1}^{N\!-\!\tN}\;
\frac{1}{4}\,b_{A}u_{A}\;\ra.
\end{equation}

To understand these formulae, let us first examine the gaugino
mass in a theory with just one dilaton
\cite{versus}\cite{dilaton}\cite{modular}. In this case the
gaugino mass is as follows \cite{versus},
\begin{equation} \label{mag3}
m_{\lambda}\,=\,
\la\,\frac{1+b\l}{b\l}\,\ra\la\,\frac{1+f-\l\df}
{1+f}\,\ra m_{\tilde{G}}.
\end{equation}
All scalar fields are evaluated at the minimum
of the potential,
\begin{equation} \label{mag2}
\cV\,=\,\frac{1}{16\l^{2}}\lbr\,
(1+f-\l\df)(1+b\l)^{2}\,-\,3b^{2}\l^{2}\,
\rbr \bar{u}u.
\end{equation}
{}From this we see that any vacuum
with zero cosmological constant satisfies $\langle 1+f-\l
\df\rangle\,=\,3b^{2}\langle\l^{2}\rangle\,+\,\CO(b^{3})$.
Since $\langle 1+f\rangle$ and $\langle \l\rangle$ are
numbers of order one, equation (\ref{mag3}) implies that
$m_{\lambda}$ is smaller than $m_{\tilde{G}}$ by a factor
of $b$.

Let us compare this to a model in which
the visible and hidden sectors couple to different
dilaton-like fields. The scalar potential of
(\ref{cond40}) receives contributions from both sectors,
proportional to $\langle 1+\tf^{\tA}-\tl_{\tA}\dtf^{\tA}\rangle$
and $\langle 1+f^{A}-\l_{A}\df^{A}\rangle$. However,
to leading order, only $\langle 1+\tf^{\tA}-\tl_{\tA}
\dtf^{\tA} \rangle$ contributes to the gaugino masses.
As above, minimizing the potential forces $\langle
1+f^{A}-\l_{A}\df^{A}\rangle$ to be $\CO(b_{A}^{2})$,
but it does not constrain $\langle 1+\tf^{\tA}-\tl_{
\tA}\dtf^{\tA}\rangle$. This suggests that the gaugino
masses can indeed be as large as
the gravitino mass.

To see this explicitly, let us consider an example with
just two types of D-branes ($\tN$=1 and $N$=2). Each
has its own dilaton-like field, $\tl_{\tA}$
and $\l_{A}$. The scalar potential follows from
(\ref{cond40}), with no summation over the indices $A$
and $\tA$:
\begin{equation} \label{mag4}
\cV\,=\,\frac{1}{16\l_{A}^{2}}\lbr\;
(1+f^{A}-\l_{A}\df^{A})(1+b_{A}\l_{A})^{2}
\,+\,b_{A}^{2}\l_{A}^{2}
\[\,(1+\tf^{\tA}-\tl_{\tA}\dtf^{\tA})\,-\,3\,\]
\;\rbr \bar{u}_{A}u_{A}.
\end{equation}
The conditions for a nontrivial vacuum with vanishing
cosmological constant are
\begin{eqnarray}
& &\la\,\frac{\pp}{\pp\tl_{\tA}}
\(1+\tf^{\tA}-\tl_{\tA}\dtf^{\tA}\)\,\ra\,=\,0,
\label{vacuum1} \nonumber \\
& &\la\,\frac{\pp}{\pp\l_{A}}\(1+f^{A}-\l_{A}\df^{A}\)\,\ra\,=\,
2b_{A}^{2}\langle\l_{A}
\rangle\[\,3\,-\,\langle 1+\tf^{\tA}-\tl_{\tA}\dtf^{\tA}\rangle\,\]
\,+\,\CO(b_{A}^{3}),
\label{vacuum2} \nonumber \\
& &\langle 1+f^{A}-\l_{A}\df^{A}\rangle\,=\,
b_{A}^{2}\langle\l_{A}^{2}
\rangle\[\,3\,-\,\langle 1+\tf^{\tA}-\tl_{\tA}\dtf^{\tA}\rangle\,\]
\,+\,\CO(b_{A}^{3}),
\label{vacuum3}
\end{eqnarray}
where all $\CO(b_{A}^{3})$ terms are suppressed. These equations
have a consistent solution if $\langle 1+f^{A}-\l_{A}
\df^{A}\rangle=\CO(b_{A}^{2})$ and $\langle 1+\tf^{\tA}-\tl_{\tA}
\dtf^{\tA}\rangle$ is $\CO(1)$. Since this latter term fixes
the visible-sector gaugino mass, we expect $m_{\lambda}^{\tA}$
to be of order $m_{\tG}$.

To actually compute the gaugino mass, we need to specify
the functions $f^A$ and $\tf^{\tA}$. (The functions
$g_{A}$ and $\tg_{\tA}$ are determined via (\ref{cond5})).
We choose them to stabilize the runaway vacuum
typically associated with dilaton-like
fields.\footnote{See \cite{Dine99}\cite{Pol} for recent
reviews.} The general procedure is described in
\cite{dilaton}\cite{Banks94}. For now,
we consider the following simple choice,\footnote{This
choice is motivated by the
Type I string theory discussed in \cite{Eva}.}
\begin{equation} \label{vacuum4}
f_{A}=\cA\cdot e^{-\cB/\l_{A}},\;\;\;\,\,\,
\tf_{\tA}=\tcA\cdot e^{-\tcB/\tl_{\tA}},\;\;\;\,\,\,
\cA,\cB,\tcA,\tcB > 0.
\end{equation}
Substituting (\ref{vacuum4}) into (\ref{vacuum3}), we
find that the potential has an extremum, located at
\begin{eqnarray}
\langle\,\tl_{\tA}\,\rangle\,&=&\,\frac{\tcB}{2},
\label{vacuum5} \nonumber\\
\langle\,\l_{A}\,\rangle\,&=&\,\frac{\cB}{2}\,\,+\,
\,\CO(b_{A}^{2}),
\label{vacuum6} \nonumber\\[1mm]
\cA\,&=&\,e^{2}\,\,+\,\,\CO(b_{A}^{2}),
\label{vacuum7}
\end{eqnarray}
for any value of $\tcA$.  It is, in fact, a global minimum
of the potential with vanishing cosmological constant, as
illustrated in Figures 1 and 2.

\begin{figure}
\centerline{
\epsfxsize=11cm
\epsfysize=9cm
\epsfbox{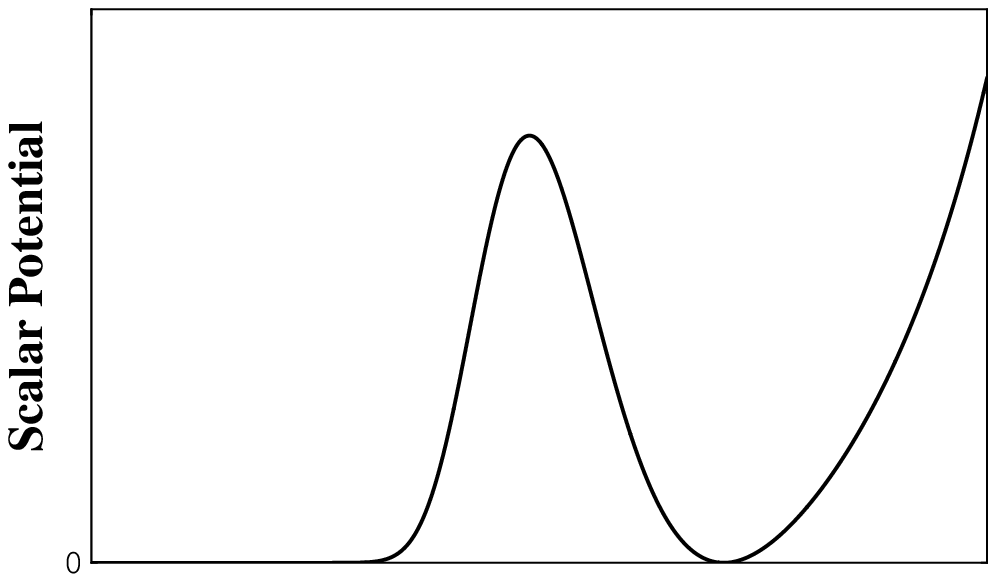}}
\caption{The scalar potential, $\cV$, plotted versus
$\l_{A}$ for fixed $\tl_{\tA}=\langle\tl_{\tA}\rangle$.}
\end{figure}
\begin{figure}
\centerline{
\epsfxsize=11cm
\epsfysize=9cm
\epsfbox{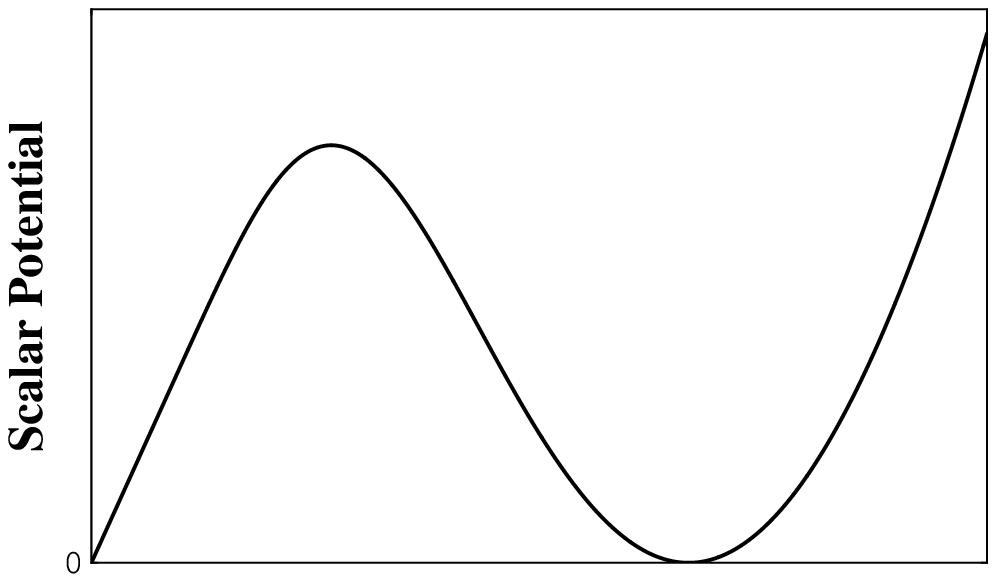}}
\caption{The scalar potential, $\cV$, plotted versus
$\tl_{\tA}$ for fixed $\l_{A}=\langle\l_{A}\rangle$.}
\end{figure}

Let us now evaluate the gaugino mass in this vacuum.
We find
\begin{equation} \label{vacuum8}
m^{\tA}_{\lambda}\,=\,
\frac{\,e^{2}-\tcA\,}{\,e^{2}+\tcA\,}\,m_{\tilde{G}},
\end{equation}
where
\vspace{-0.7cm}
\begin{eqnarray}
0.2\,m_{\tG}\;<\;&m^{\tA}_{\lambda}&\;<\;0.8\,m_{\tG}\;\;\;\;\;
\mbox{for}
\nonumber \\
4.9\;>\;&\tcA&\;>\;0.8\,.
\label{vacuum9}
\end{eqnarray}
We see that, for reasonable values of $\tcA$, the gaugino
mass is of order the gravitino mass.

\section{Conclusion} \label{conclusion} \setcounter{equation}{0}

The presence of multiple dilaton-like moduli fields is a very
important feature of ${\cal N}$=1 D=4 Type I string models.
The extra dilaton-like fields can change the resulting
phenomenology in many ways. In this paper we examined their
effect on supersymmetry breaking by hidden-sector gaugino
condensation.

We studied a scenario in which different dilaton-like fields
couple to the hidden and visible sectors. We assumed
that supersymmetry is broken by gaugino condensation
in the hidden sector. We found that the visible-sector
gaugino masses can be as large as gravitino mass because
of the extra dilaton-like fields. Our results stand in
contrast to the usual heterotic string phenomenology,
where the gaugino masses are typically much smaller.

We would like to thank Mary K.~Gaillard and Gary
Shiu for useful discussions. This
work was supported in part by NSF grant No. PHY-9404057.

\section*{Appendix}
\setcounter{section}{0}
\renewcommand{\thesection}{\Alph{section}}
\renewcommand{\theequation}{\Alph{section}.\arabic{equation}}

\section{Linear--Chiral Duality} \setcounter{equation}{0}

In this Appendix we prove that a supergravity
model with $N$ linear supermultiplets is dual
to another supergravity model with $N$ chiral
supermultiplets.\footnote{This duality
was briefly discussed in \cite{Binetruy91}.}
We will show that both models can be obtained
from the following Lagrangian,
\begin{equation} \label{model7}
\Lag\,=\,\superint\,E\,\lbr\; F(V_{1},\cdots,V_{N})\,+\,
\sum_{A=1}^{N}(S_{A}+\bar{S}_{A})(\Omega_{A}-V_{A}) \;\rbr
\end{equation}
with
\begin{equation}
K\,=\,K(V_{1},\cdots,V_{N}) \nonumber \\
\end{equation}
and
\begin{equation} \label{model8}
-3\,+\,\sum_{A=1}^{N}V_{A}\frac{\pp K}{\pp V_{A}} \,=\,
F\,-\,\sum_{A=1}^{N}V_{A}\frac{\pp F}{\pp V_{A}}.
\end{equation}
In these expressions, the $V_A$'s are {\em unconstrained}
real superfields and the $S_A$'s are ordinary chiral
supermultiplets, for $A=1,\cdots,N$.  The field $\Omega_{A}$
is a Chern-Simons superform; it obeys
\begin{equation}
-(\bar{\Diff}^{2}-8R)\Omega_{A}\,=\,
{\cal W}_{A}{\cal W}_{A},
\end{equation}
for $A=1,\cdots,N$.

Let us first integrate out the $S_{A}$'s in $\Lag$. Their
equations of motion are as follows,
\begin{equation} \label{model9}
-(\bar{\Diff}^{2}-8R)V_{A}\,=\,
{\cal W}_{A}{\cal W}_{A}.
\end{equation}
This equation can be used to eliminate the second
term in (\ref{model7}), reducing $\Lag$ to a
model with $N$ linear supermultiplets.  Note that
the Einstein gravity term is canonically normalized
because of (\ref{model8}).

Let us now return to $\Lag$ and integrate out the
$V_{A}$'s. Their equations of motion are as follows,
\begin{equation} \label{model10}
S_{A}+\bar{S}_{A}\,=\,\frac{\pp F}{\pp V_{A}}\,-\,
\frac{1}{3}\frac{\pp K}{\pp V_{A}}
\lbr F\,-\,\sum_{B=1}^{N}V_{B}(S_{B}+\bar{S}_{B}) \rbr.
\end{equation}
If we multiply (\ref{model10}) by $V_{A}$ and sum over
$A=1,\cdots,N$, we find\footnote{Equation
(\ref{duality1}) is obtained by assuming $\sum_{A=1}^{N}
V_{A}\frac{\pp K} {\pp V_{A}}\,\neq\,3$, as is true for
the models considered in this paper.}
\begin{equation} \label{duality1}
F(V_{1},\cdots,V_{N})\,-\,
\sum_{A=1}^{N}V_{A}(S_{A}+\bar{S}_{A})\,=\,-3.
\end{equation}
It is now trivial to use (\ref{duality1}) to rewrite
(\ref{model10}) in a very simple form:
\begin{equation} \label{duality2}
S_{A}+\bar{S}_{A}\,=\,\frac{\pp \(K+F\)}{\pp V_{A}}.
\end{equation}
Using these relations, we find
\begin{equation} \label{duality3}
\Lag\,=\,-3 \superint\,E\,\ +\
\Bigg\{\,\sum_{A=1}^{N}\frac{1}{8}\superint\,
         \frac{E}{R}\,
         S_{A}({\cal W}^{\alpha}{\cal W}_{\alpha})_{A}\,+\,
\mbox{h.c.}\Bigg\},
\end{equation}
with
\begin{equation}
K\,=\,{K}(S_{1}+\bar{S}_{1},\cdots,S_{N}+\bar{S}_{N}).
\end{equation}
This completes the proof
of linear--chiral duality.

\section{Component-Field Lagrangian}
\setcounter{equation}{0}

In this Appendix, we compute the necessary elements of the
component-field Lagrangian corresponding to the superfield
Lagrangian (\ref{cond2})--(\ref{cond5}).  We use
the chiral density method \cite{dilaton}\cite{Adamietz}.

We start by enumerating the definitions of bosonic component
fields.  In the hidden sector, we have
\begin{eqnarray}
\l_{A}\,&=&\,V_{A}\lowest,   \label{cond7}\nonumber\\
2\sigma^{m}_{\alpha\dot{\alpha}}B_{m}^{A}\,-\,
\frac{4}{3}\l^{A}\sigma^{a}_{\alpha\dot{\alpha}} b_{a}\,&=&\,
[\,\Diff_{\alpha},\Diff_{\dot{\alpha}}\,]V^{A}\lowest,
\label{cond8}\nonumber\\
u_{A}\,&=&\,U_{A}\lowest\,\equiv\,-(\bar{\Diff}^{2}-8R)V_{A}\lowest,
\label{cond9}\nonumber\\
-4F_{A}\,&=&\,-\Diff^{2}(\bar{\Diff}^{2}-8R)V_{A}\lowest.
\label{cond11}
\end{eqnarray}
In these expressions, the $\l_{A}$ are dilaton-like scalar
fields, and the $B_{m}^{A}$ are axionic degrees of freedom
in the same supermultiplets.  The fields $u_{A}$ are the
gaugino condensate fields of the hidden sector.

The visible-sector fields are defined in a similar
way,\footnote{We include the gaugino fields for
the reader's convenience.}
\begin{eqnarray}
\tl_{\tA}\,&=&\,\tV_{\tA}\lowest, \label{cond15}\nonumber\\
2\sigma^{m}_{\alpha\dot{\alpha}}\tB_{m}^{\tA}\,-\,
\frac{4}{3}\tl^{\tA}\sigma^{a}_{\alpha\dot{\alpha}} b_{a}
\,+\,2\Tr(\lambda_{\alpha}^{\tA}\bar{\lambda}_{\dot{\alpha}}^{\tA})
\,&=&\,
[\,\Diff_{\alpha},\Diff_{\dot{\alpha}}\,]\tV^{\tA}\lowest,
\label{cond16}\nonumber\\
-\Tr(\lambda^{\tA}\lambda^{\tA})\,&=&\,
-(\bar{\Diff}^{2}-8R)\tV_{\tA}\lowest,
\label{cond17}\nonumber\\
\vspace{0.8cm}
8i\Tr(\lambda^{\tA}\sigma^{m}\Diff_{m}\bar{\lambda}^{\tA})\,+\,
4\Tr(\lambda^{\tA}\lambda^{\tA})\bar{M}\,& & \nonumber \\
+\,2\Tr(\tF^{mn}_{\tA}\tF_{mn}^{\tA})
\,+\,i\epsilon^{mnpq}\Tr(\tF_{mn}^{\tA}\tF_{pq}^{\tA})\,-\,
4\Tr(\tD^{\tA}\tD^{\tA})
\,&=&\,-\Diff^{2}(\bar{\Diff}^{2}-8R)\tV_{\tA}\lowest,
\;\;\;\;\;\;\;\;\;\;
\label{cond19}
\end{eqnarray}
where
\begin{eqnarray}
-i\lambda_{\alpha}^{\tA}\,&=&\,{\cal W}_{\alpha}^{\tA}\lowest,
\;\;\;\;\;
i\bar{\lambda}_{\dot{\alpha}}^{\tA}\,=\,
{\cal W}_{\dot{\alpha}}^{\tA}\lowest,\label{cond21}\nonumber\\
-2\tD^{\tA}\,&=&\,\Diff^{\alpha}{\cal W}_{\alpha}^{\tA}\lowest\,=\,
\Diff_{\dot{\alpha}}{\cal W}^{\dot{\alpha}}_{\tA}\lowest,
\label{cond22}\nonumber\\
\tB^{m}_{\tA}\,&=&\,\frac{1}{2}\epsilon^{mnpq}
\{\,\pp_{q}\tilde{b}_{pn}^{\tA}\,+\,
\frac{1}{6}\Tr(\ta_{[q}\pp_{p}\ta_{n]}\,-\,
\frac{2i}{3}\ta_{[q}\ta_{p}\ta_{n]})^{\tA}\,\}. \label{cond23}
\end{eqnarray}
In these expressions, the $\tl_{\tA}$ are dilaton-like
scalar fields.  The $\tB_{m}^{\tA}$ are dual field
strengths of the antisymmetric tensors $\tilde{b}_{p
q}^{\tA}$.  The $\tF_{mn}^{\tA}$ are the Yang-Mills
field strengths, while the $\ta_{m}^{\tA}$ are the corresponding
gauge fields.  The fields $M$, $\bar{M}$, and $b_{a}$ are the
auxiliary fields of the supergravity multiplet \cite{Wess}.
The bosonic components of the $\Phi_{i}$ are $\phi_{i}=
\Phi_{i}\lowest$ and $-4F_{i}=\Diff^{2}\Phi_{i}\lowest$.

Using these definitions, we find the following bosonic
component-field Lagrangian:
\begin{eqnarray} \label{cond29}
\frac{1}{\sqrt{-g\,}}\Lag_{B}\,
&=&\,-\,\frac{1}{4}{\cal R}
\,-\,\sum_{\tA=1}^{\tN}
\frac{1}{8\tl_{\tA}^{2}}\(1+\tl_{\tA}\dtg^{\tA}\)
\nabla^{m}\!\tl_{\tA}\,\nabla_{\!m}\!\tl_{\tA}
\nonumber\\
& &\,+\,\sum_{\tA=1}^{\tN}
\frac{1}{8\tl_{\tA}^{2}}\(1+\tl_{\tA}\dtg^{\tA}\)
\tB^{m}_{\tA}\!\tB_{m}^{\tA}
\,-\,\sum_{\tA=1}^{\tN}
\frac{1+\tf_{\tA}}{16\tl_{\tA}}\;\Tr\!\(\tF^{mn}_{\tA}\tF_{mn}^{\tA}\)
\nonumber\\
& &\,-\,\sum_{A=1}^{N\!-\!\tN}\frac{1}{8\l_{A}^{2}}\(1+\l_{A}\dg^{A}\)
\nabla^{m}\!\l_{A}\,\nabla_{\!m}\!\l_{A}
\,+\,\sum_{A=1}^{N\!-\!\tN}\frac{1}{8\l_{A}^{2}}\(1+\l_{A}\dg^{A}\)
B^{m}_{A}\!B_{m}^{A}
\nonumber\\
& &\,+\,\sum_{A=1}^{N\!-\!\tN}\frac{i}{4}b_{A}
B^{m}_{A}\nabla_{\!m}\!\ln\!\(\frac{\bar{u}_{A}}{u_{A}}\)
\,-\,\frac{1}{2}\sum_{i,j}G_{i\bar{j}}
\nabla^{m}\phi_{i}\,\nabla_{\!m}\bar{\phi}_{j}
\nonumber\\
& &\,-\,\frac{1}{18}\(\,N\,-\,3\,+\,
\sum_{\tA=1}^{\tN}\tl_{\tA}\dtg^{\tA}
\,+\,\sum_{A=1}^{N\!-\!\tN}\l_{A}\dg^{A}\,\)b^{a}b_{a}
\,+\,\sum_{\tA=1}^{\tN}
\frac{1+\tf_{\tA}}{8\tl_{\tA}}\;\Tr\!\(\tD^{\tA}\tD^{\tA}\)
\nonumber\\
& &\,+\,\frac{1}{2}\sum_{i,j}G_{i\bar{j}}F_{i}\bar{F}_{j}
\,-\,\frac{1}{4}\(\sum_{A=1}^{N\!-\!\tN}b_{A}u_{A}\)
\!\ \!\sum_{i}G_{i}F_{i}
\nonumber \\
& &\,+\,\sum_{A=1}^{N\!-\!\tN}
\frac{1}{8\l_{A}}\lbr\,1\,+\,f_{A}\,+\,b_{A}\l_{A}\ln\!\(e^{\!-K}
\bar{u}_{A}u_{A}\)\,+\,2b_{A}\l_{A}\,\rbr F_{A}
\nonumber\\
& &\,+\,\frac{1}{18}\lbr\,N\,-\,3\,+\,
\sum_{\tA=1}^{\tN}\(\tf^{\tA}-\tl_{\tA}\dtf^{\tA}\)
\,+\,\sum_{A=1}^{N\!-\!\tN}\(f^{A}-\l_{A}\df^{A}\)\,\rbr\bar{M}\!M
\nonumber\\
& &\,-\,\sum_{A=1}^{N\!-\!\tN}
\frac{1}{8\l_{A}}\lbr\,1\,+\,f_{A}\,+\,b_{A}\l_{A}\ln\!\(e^{\!-K}
\bar{u}_{A}u_{A}\)\,\rbr u_{A}\bar{M}
\nonumber\\
& &\,-\,\frac{1}{12}\(\sum_{B=1}^{N\!-\!\tN}b_{B}\bar{u}_{B}\)\!
\lbr\,N\,+\,\sum_{\tA=1}^{\tN}\(\tf^{\tA}-\tl_{\tA}\dtf^{\tA}\)
\,+\,\sum_{A=1}^{N\!-\!\tN}\(f^{A}-\l_{A}\df^{A}\)\,\rbr M
\nonumber\\
& &\,-\,\sum_{A=1}^{N\!-\!\tN}
\frac{1}{32\l_{A}^{2}}\(1+f^{A}-\l_{A}\df^{A}\)
\bar{u}_{A}u_{A}
\nonumber\\
& &\,-\,\frac{1}{16}\(\sum_{B=1}^{N\!-\!\tN}b_{B}\bar{u}_{B}\)\!
\lbr\,\sum_{A=1}^{N\!-\!\tN}
\frac{\(1+f^{A}-\l_{A}\df^{A}\)}{\l_{A}}u_{A}\,\rbr
\nonumber\\
& &\,+\,\mbox{h.c.}
\end{eqnarray}
We also find the following kinetic and mass terms for the gauginos and
gravitino,
\begin{eqnarray} \label{cond30}
\frac{1}{\sqrt{-g\,}}\Lag_{m_{\lambda}}\!\!\!\!
&=&\!\!\!\!-\sum_{\tA=1}^{\tN}\frac{1+\tf_{\tA}}{4\tl_{\tA}}
\Tr\!\(i\lambda^{\tA}\sigma^{m}\nabla_{\! m}\bar{\lambda}^{\tA}\)
\nonumber\\
&&\!\!\!\!
\,+\,\frac{1}{16}\(\sum_{A=1}^{N\!-\!\tN}b_{A}\bar{u}_{A}\)\!\!
\lbr\sum_{\tA=1}^{\tN}
\frac{\(1+\tf^{\tA}-\tl_{\tA}\dtf^{\tA}\)}{\tl_{\tA}}
\Tr\!\(\lambda^{\tA}\lambda^{\tA}\)\rbr\,+\,\mbox{h.c.}
\nonumber\\[3mm]
\frac{1}{\sqrt{-g\,}}\Lag_{m_{\tilde{G}}}\!\!\!\!
&\,=\,&\!\!\!\!\frac{1}{2}\epsilon^{mnpq}
\bar{\psi}_{m}\bar{\sigma}_{n}\!\nabla_{\!p}\psi_{q}\nonumber\\
&&\!\!\!\!\,-\,\sum_{A=1}^{N\!-\!\tN}\frac{1}{8\l_{A}}
\lbr 1+f_{A}+b_{A}\l_{A}\ln\!\(e^{\!-K}\bar{u}_{A}u_{A}\)
\rbr\bar{u}_{A}(\psi_{m}\sigma^{mn}\psi_{n})\,+\,\mbox{h.c.}
\nonumber\\
\end{eqnarray}

To extract the potential (\ref{cond40}), we must eliminate
the auxiliary fields.  We will not do that here, except
to note that the equations of motion for the auxiliary fields
$(F_{A}+\bar{F}_{A})$ are
\begin{equation} \label{cond33}
1\,+\,f_{A}\,+\,b_{A}\l_{A}\ln\!\(e^{\!-K}
\bar{u}_{A}u_{A}\)\,+\,2b_{A}\l_{A}\,=\,0.
\end{equation}
This fixes the modulus of the condensate field $u_{A}$ to be
\begin{equation} \label{cond34}
u_A \bar{u}_{\bar{A}}\,=\,\exp \left[\,\bigg(K-2 \bigg)\,-\,
\left( {1+f_{A}\over b_{A}\l_{A}} \right) \right] .
\end{equation}
The modulus has the correct dependence on the gauge group
and gauge coupling, as expected from the usual renormalization
group arguments \cite{versus}\cite{modular}.

\section{Gaugino Masses in the Chiral Supermultiplet Formulation}
\setcounter{equation}{0}

In this Appendix we compute the gaugino masses using the
\cs formulation.  By linear--chiral duality, the result must
be identical to the one obtained in Section \ref{gaugino}.

Following Appendix A, it is straightforward to write down the
\cs formulation of the model defined by (\ref{cond2})--(\ref{cond5}).
The real superfields $\tV_{\tA}$ and $V_{A}$ dualize to chiral
superfields\footnote{It can be shown that
the linear--chiral duality established in Appendix A also applies
to the effective theory description of Veneziano and Yankielowicz
\cite{Vene}. In particular, the linear--chiral duality relations
remain the same.}  $\tS_{\tA}$ and $S_{A}$.
The linear--chiral duality relations include
(\ref{model9}) and
\begin{equation} \label{origin10}
\tS_{\tA}+\bar{\tS}_{\tA}\,=\,
\frac{1\,+\,\tf_{\tA}(\tV_{\tA})}{\tV_{\tA}}.
\end{equation}
This gives rise to the following identities:
\begin{eqnarray}
\ts_{\tA}+\bar{\ts}_{\tA}\,&=&\,\frac{1\,+\,\tf_{\tA}}{\tl_{\tA}},
\label{origin11} \nonumber\\
\langle\,\bar{F}_{\bar{\tS}_{\tA}}\,\rangle\,&=&\,-\,
\frac{\,\la\,M\,\ra\,}{3}\
\la\,\frac{1+\tf^{\tA}-\tl_{\tA}\dtf^{\tA}}{\tl_{\tA}}\,\ra.
\label{origin12}
\end{eqnarray}
The first is the $\theta$=$\bar{\theta}$=0 component
of (\ref{origin10}).  The second is obtained by acting on both
sides of (\ref{origin10}) with the operator $\bar{\Diff}^{2}$,
and then taking the vacuum expectation value of the lowest
component.

The expression for the gaugino masses is standard,
\begin{equation} \label{origin9}
m^{\tA}_{\lambda}\,=\,-\,
\frac{\,\langle\,\bar{F}_{\bar{\tS}_{\tA}}\,\rangle\,}
{\,\langle\,\ts_{\tA}+\bar{\ts}_{\tA}\,\rangle\,},
\end{equation}
where $\ts_{\tA}=\tS_{\tA}\lowest$ and $F_{\tS_{\tA}}=
- \frac{1}{4}\Diff^{2}\tS_{\tA}\lowest$. Using
(\ref{origin11}) and $\la M\ra\,=\,3m_{\tilde{G}}$,
we find a final result for the gaugino masses that
is identical to (\ref{cond43}), obtained from the
\ls formalism.

\end{document}